\documentclass[runningheads]{llncs}

\usepackage{graphicx}
\usepackage{ctable}
\usepackage{multirow}
\usepackage[caption=false]{subfig}

\begin{document}

\title{Recommendations in a Multi-Domain Setting: Adapting for Customization, Scalability and Real-Time Performance}

\titlerunning{Recommendations in a Multi-Domain Setting}

\author{Emanuel Lacic \inst{1} \and Dominik Kowald \inst{1,2}}
\authorrunning{E. Lacic, and D. Kowald}

\institute{Know-Center GmbH, Graz, Austria \\\email{{elacic,dkowald}@know-center.at} \and
Graz University of Technology, Graz, Austria}

\maketitle

\begin{abstract}
In this industry talk at ECIR'2022, we illustrate how to build a modern recommender system that can serve recommendations in real-time for a diverse set of application domains. Specifically, we present our system architecture that utilizes popular recommendation algorithms from the literature such as Collaborative Filtering, Content-based Filtering as well as various neural embedding approaches (e.g., Doc2Vec, Autoencoders, etc.). We showcase the applicability of our system architecture using two real-world use-cases, namely providing recommendations for the domains of (i) job marketplaces, and (ii) entrepreneurial start-up founding. We strongly believe that our experiences from both research- and industry-oriented settings should be of interest for practitioners in the field of real-time multi-domain recommender systems.

\end{abstract}

\section{Recommendations in a Multi-Domain Setting}
\label{s:abs}

Recommender systems~\cite{resnick1997recommender} have gained a tremendous increase in popularity in recent years for many industry practitioners. Early recommender systems often considered only user-item interactions, but nowadays, many application domains can leverage different contextual sources like textual meta-data \cite{yao2015collaborative}, images \cite{kang2017visually} or implicitly arising graph structures \cite{Jamali2010Trust}. Furthermore, practitioners who build modern recommender systems need to address the scalability and real-time demand when providing recommendations in an online setting~\cite{Chandramouli2011,davidson2010youtube,lacic2014towards,shani2005mdp}, since there is usually a trade-off between accuracy and runtime performance. When put into production, different challenges need to be addressed in order to continuously maintain the stability and health of a recommender system.  A distributed architecture which is guided by design principles like providing service isolation, supporting data heterogeneity, allowing for algorithmic customization as well as ensuring fault tolerance is thus a necessity. 

\begin{figure}[t]
\centering

      \includegraphics[width=.6\textwidth]{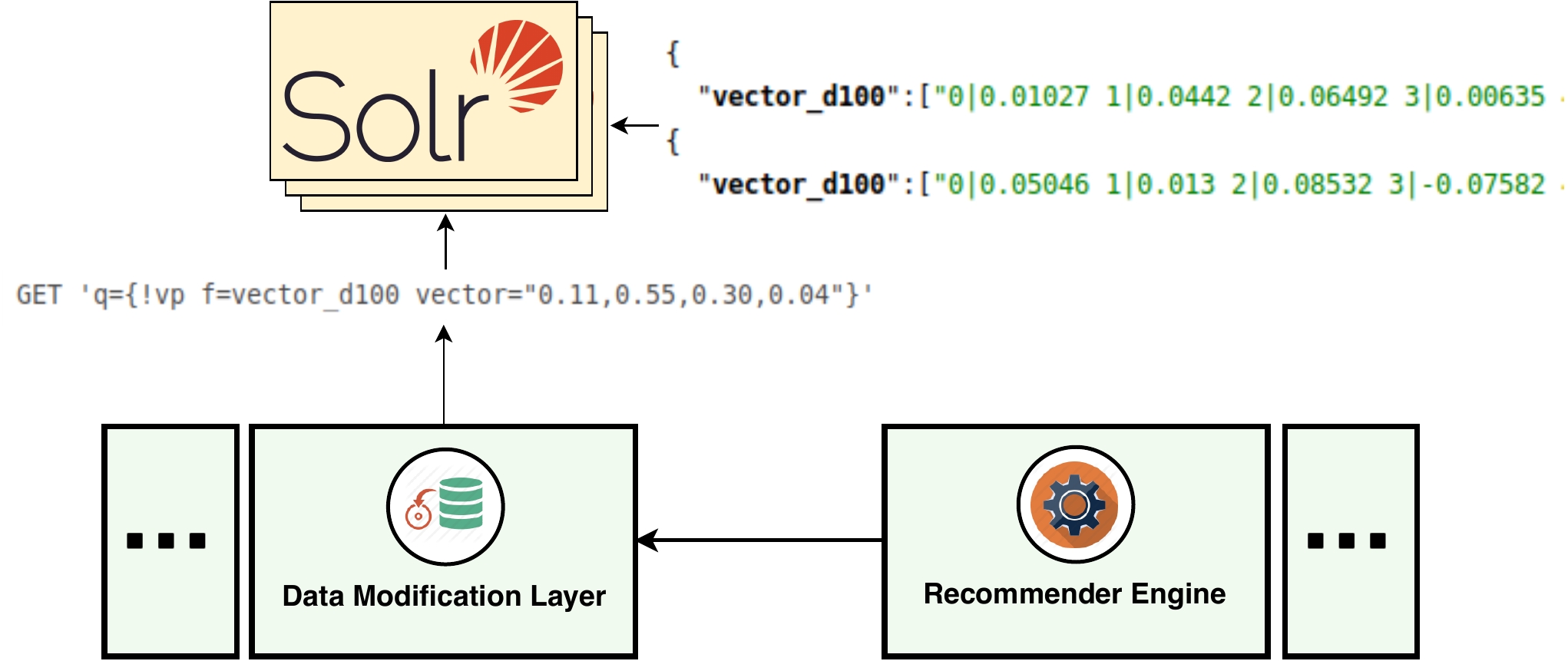}

   \caption{Fast similarity calculations that rely on latent embeddings can be achieved, for instance, by using search engine technology like Apache Solr.}
   \label{fig:solr}
\end{figure}

\vspace{2mm}
\noindent
In this industry talk, we will show how to build a modern recommender system that can serve recommendations in real-time for a diverse set of application domains. We will share our experiences that we gained in both research-oriented (e.g., Horizon 2020 or FP7~\cite{kowald2013social}) and industry-oriented (e.g., online~\cite{lacic2013utilizing} or offline~\cite{lacic2015tackling} marketplaces) projects on how we build hybrid models based on a microservice architecture. This architecture utilizes popular algorithms from the literature such as Collaborative Filtering, Content-based Filtering as well as various neural embedding approaches (e.g., Doc2Vec, Autoencoders, etc.). As depicted in Figure \ref{fig:solr}, we will further show how we adapt our architecture to calculate relevant recommendations in real-time (i.e., after a recommendation is requested), since in many cases individual requests may be targeted for user sessions that are short-lived and context-dependent.

\begin{figure}[t!]
\centering

   \subfloat[Job Marketplace]{
      \includegraphics[width=.40\textwidth]{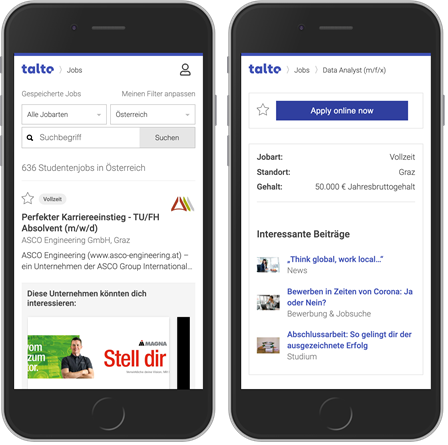}
      \label{fig:talto}
      }
   ~   
\subfloat[Co-Founder Matching]{
      \includegraphics[width=.52\textwidth]{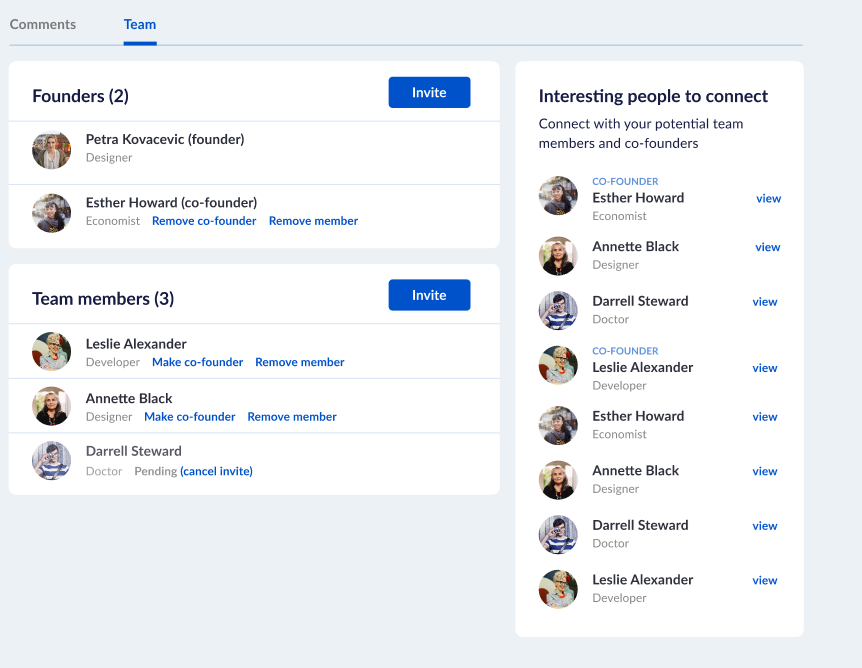}
       \label{fig:cogsteps}
      }

   \caption{In a multi-domain setting, the same system supports different personalization scenarios, e.g., (a) and (b), and relies on recommender algorithms that need to be customized based on the underlying data structure and optimization criteria of the respective domain.}
   \label{fig:domain}
\end{figure}

\vspace{2mm}
\noindent
To showcase the applicability of such an approach, we will specifically focus on and present two real-world use-cases, namely providing recommendations for the domains of (i) job marketplaces as seen in Figure \ref{fig:talto}, and (ii)  entrepreneurial start-up founding shown in Figure \ref{fig:cogsteps}. For the former, we tackle the problem of finding the right job for university students \cite{reiter2020heterogeneous} by guiding the students toward different types of entities that are related to their career, i.e., job postings, company profiles, and career-related articles. Here, for instance, we find that the online performance of the utilized approach also depends on the location context where the recommendations are displayed. For the latter, we will present how a recommender system can support academic entrepreneurs who want to go through the process of building a start-up from an initial innovation idea. In such a setting, a recommender system needs to suggest relevant experts that can provide feedback to an innovation idea, support potential co-founder and team member matching, allow accelerators, incubators, and innovation hubs to discover these innovations as well as continuously provide relevant education materials until the innovation idea has become mature enough in order to form a start-up. By adapting a recommender system for such diverse personalization scenarios, we observe that a dynamic customization of the utilized recommender algorithms with respect to the underlying data structures is of key importance.

\vspace{2mm}
\noindent
Taken together, we strongly believe that our experiences from both research- and industry-oriented projects should be of interest for the ECIR audience, especially for practitioners in the field of real-time multi-domain recommender systems.

\section{Company Portrait and Speakers}
\label{s:cp}

With more than 20 years of experience in cutting-edge research, Know-Center GmbH is Austria's leading research center for Data-Driven Business and Big Data Analytics. In its role as an innovation hub between science and industry, the Know-Center as a non-profit company offers application-oriented research in cooperation with academic institutions and partners from industry. The scientific strategy of the Know-Center is to combine approaches of Big Data Analytics with Human-Centered Computing to create cognitive computing systems that enable humans to use huge amounts of data. With over 130 employees, Know-Center has extensive experience in national as well as international collaborative R\&D projects in Big Data, Machine Learning and Artificial Intelligence. 


\vspace{2mm}
\noindent
\emph{Speaker 1:} Emanuel Lacic, MSc. is Operations Area Manager of the Social Computing area at the Know-Center. He is a former Marshall Plan fellow and has been working as a visiting researcher at the Computer Science department of the University of California, Los Angeles. His main research interests are in the fields of Recommender Systems, Deep Learning, as well as Social Network Analysis.

\vspace{2mm}
\noindent
\emph{Speaker 2:} Dr. Dominik Kowald is Research Area Manager of the Social Computing area at the Know-Center and senior researcher at Graz University of Technology. He is also task lead in the H2020 AI4EU and TRUSTS projects. He is review editor of Frontiers in Big Data – Recommender Systems section, and his research interests are in Recommender Systems, Privacy, Fairness, and Biases in algorithms.

\vspace{2mm}
\noindent
\textbf{Acknowledgements.} This work was funded by the H2020 projects TRUSTS (GA: 871481), TRIPLE (GA: 863420), and the Erasmus+ project COGSTEPS.

\bibliographystyle{splncs04}
\bibliography{bib}

\end{document}